\documentclass[journal]{IEEEtran}
\usepackage{amsmath,amsfonts}
\usepackage{array}
\usepackage[justification=justified,font=small]{caption}	
\usepackage[labelformat=simple]{subcaption}

\usepackage{textcomp}
\usepackage{stfloats}
\usepackage{url}
\usepackage{verbatim}
\usepackage{graphicx}
\usepackage{cite}
\usepackage{setspace}
\usepackage{epsfig}  
\usepackage{graphicx}  
\usepackage{amssymb}  
\usepackage{lscape}  
\usepackage{color}
\usepackage{mathrsfs}
\usepackage{mathtools}
\usepackage{amsthm}
\usepackage{array}
\usepackage{booktabs}
\usepackage{cases}
\usepackage{bm}
\usepackage{textcomp}
\newcommand{\textapprox}{\raisebox{0.5ex}{\texttildelow}}

\begin{document}

\title{Maxwell-Schr\"{o}dinger Modeling of Superconducting Qubits Coupled to Transmission Line Networks}

\author{Thomas E. Roth~\IEEEmembership{Member,~IEEE},
	and Samuel T. Elkin~\IEEEmembership{Member,~IEEE}
\thanks{Manuscript received XXXX XX, 2023; revised XXXX XX, 2023. \\
\indent This work was funded by a startup fund at Purdue University. \textit{(Corresponding author: Thomas E. Roth.)}\\
\indent Thomas E. Roth is with the Elmore Family School of Electrical and Computer Engineering, Purdue University, West Lafayette, IN 47907 USA and the Purdue Quantum Science and Engineering Institute, West Lafayette, IN 47907 USA (e-mail: rothte@purdue.edu).\\
\indent Samuel T. Elkin was with the Elmore Family School of Electrical and Computer Engineering, Purdue University, West Lafayette, IN 47907 USA. He is now with Indesign, LLC., Indianapolis, IN 46216 USA.} 
}

\markboth{Journal of \LaTeX\ Class Files,~Vol.~XX, No.~X, XX~2023}%
{Shell \MakeLowercase{\textit{et al.}}: A Sample Article Using IEEEtran.cls for IEEE Journals}

\maketitle

\begin{abstract}
In superconducting circuit quantum information technologies, classical microwave pulses are applied to control and measure the qubit states. Currently, the design of these microwave pulses use simple theoretical or numerical models that do not account for the self-consistent interactions of how the qubit state modifies the applied microwave pulse. In this work, we present the formulation and finite element time domain discretization of a semiclassical Maxwell-Schr\"{o}dinger method for describing these self-consistent dynamics for the case of a superconducting qubit capacitively coupled to a general transmission line network. We validate the proposed method by characterizing key effects related to common control and measurement approaches for transmon and fluxonium qubits in systems that are amenable to theoretical analysis. Our numerical results also highlight scenarios where including the self-consistent interactions are essential. By treating the microwaves classically, our method is substantially more efficient than fully-quantum methods for the many situations where the quantum statistics of the microwaves are not needed. Further, our approach does not require any reformulations when the transmission line system is modified. In the future, our method can be used to rapidly explore broader design spaces to search for more effective control and measurement protocols for superconducting qubits.
\end{abstract}


\section{Introduction}
\IEEEPARstart{S}{uperconducting} circuit architectures, commonly referred to as circuit quantum electrodynamics (cQED) devices, are one of the leading approaches for developing quantum computers \cite{arute2019quantum,wu2021strong} and other quantum information processing technologies \cite{gu2017microwave,krantz2019quantum,blais2021circuit}. Although great progress has been made, significant improvements are still needed to reach these technologies full potential. Some of the most pressing challenges are to significantly increase the number of qubits in the devices while further improving the speed and fidelity of qubit control and measurement \cite{jurcevic2021demonstration,chen2021exponential,acharya2023suppressing}. This is necessitating the exploration of new packaging and integration strategies \cite{kosen2022building,conner2021superconducting,rosenberg20173d,wang2022hexagonal}, which is complicated due to the stringent system requirements. For example, the quantum error correcting code with the most lenient requirements can still require control fidelities on the order of 0.9999 \cite{huang2019performance}. Since qubit control and measurement is accomplished with classical microwave drives in these systems \cite{krantz2019quantum}, improved microwave engineering is a key direction for meeting these system requirements. 

To explore the design space of these emerging device approaches, higher-fidelity numerical modeling methods (e.g., full-wave methods) are becoming increasingly necessary \cite{nigg2012black,solgun2019simple,roth2021macroscopic,minev2021energy,roth2021fullSPS,roth2022full}. However, these modeling methods are still in their infancy, with existing methods often requiring tedious user-intensive procedures \cite{nigg2012black} or computationally costly eigenmode decompositions of the electromagnetic system \cite{roth2021macroscopic,minev2021energy}. These methods will significantly struggle to scale to systems with many qubits, and so it is of interest to develop efficient high-fidelity numerical modeling methods for cQED devices.

One avenue to satisfy many of these modeling needs is with self-consistent semiclassical methods that treat certain electromagnetic effects classically and the qubit dynamics quantum mechanically. Semiclassical modeling has a long history in studying large-scale atom-field interactions like those involved in lasers \cite{scully2001quantum}, but have more recently begun being used in optical frequency regimes to analyze how classical electromagnetic fields can control the state of individual quantum systems \cite{takeuchi2015maxwell,ryu2016finite,xiang2017high,chen2017unified,xiang2019quantum}. Often referred to as Maxwell-Schr\"{o}dinger methods, these approaches have thus far been overlooked for cQED systems, despite the more prominent role that classical microwave fields play in the control \textit{and} measurement of qubit states in comparison to optical technologies \cite{gu2017microwave,krantz2019quantum,blais2021circuit}.

In this work, we present a systematic formulation and general-purpose Maxwell-Schr\"{o}dinger discretization approach for self-consistent semiclassical equations of motion describing a superconducting qubit capacitively coupled to a transmission line network. We present our derivation specifically for the transmon qubit \cite{koch2007charge,roth2022transmon} to make the discussion concrete, but also include results for a fluxonium qubit \cite{manucharyan2009fluxonium} to demonstrate the general applicability of the method. By avoiding the need to perform any electromagnetic eigenmode decompositions, our self-consistent semiclassical framework is numerically scalable. Further, our discretization method can be generalized in a straightforward manner to consider arbitrary transmission line networks that would be impossible to treat theoretically or be prohibitively cumbersome to consider using typical bespoke models that must be reformulated any time the transmission line system is substantively modified \cite{peropadre2011approaching,peropadre2013scattering}. In comparison to typical master equation approaches \cite{johansson2012qutip}, our approach can be orders of magnitude more efficient when analyzing situations where the quantum statistics of the microwaves are not essential.

Preliminary results on this formulation were reported in \cite{roth2022derivation,roth2023maxwell,roth2023hybrid,roth2023finite}. This work expands on \cite{roth2022derivation,roth2023maxwell,roth2023hybrid,roth2023finite} by providing comprehensive details on the theoretical formulation, a comparison of different discretization techniques, a stability analysis of the numerical method, and extending the approach to fluxonium qubits. There are also a significant number of new numerical results, including ones that support a more detailed quantitative validation of the numerical method than previously presented.

The remainder of this work is organized in the following manner. In Section \ref{sec:formulation}, we present the derivation of a self-consistent set of semiclassical equations of motion that characterize the dynamics of a superconducting qubit capacitively coupled to a transmission line. We then describe a suitable one-dimensional Maxwell-Schr\"{o}dinger method to numerically solve the semiclassical equations of motion in Section \ref{sec:discretization}. Our approach is specifically designed to ease the transition to a full-wave three-dimensional Maxwell-Schr\"{o}dinger method in the future. We present a range of numerical results in Section \ref{sec:results} to validate our numerical method against established theoretical predictions. Finally, we discuss conclusions and directions for future work in Section \ref{sec:conclusion}.
\section{Formulation}
\label{sec:formulation}
\begin{figure}[t!]
	\centering
	\includegraphics[width=0.75\linewidth]{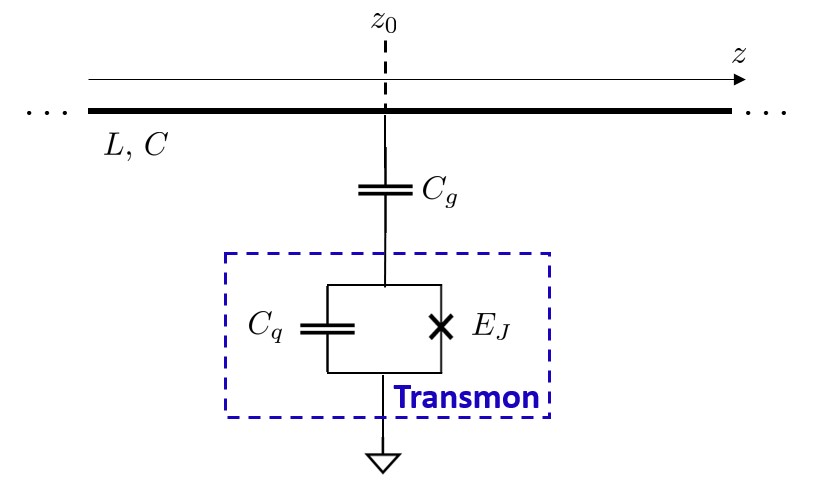}
	\caption{Schematic of a transmon qubit capacitively coupled to a transmission line with per-unit-length parameters of $L$ and $C$. A transmon qubit is composed of a capacitance $C_q$ in parallel with a Josephson junction that has Josephson energy $E_J$.}
	\label{fig:transmon-tr-line-coupling}
\end{figure}

In this section, we derive self-consistent semiclassical equations of motion for a transmon capacitively coupled to a transmission line, as shown in Fig. \ref{fig:transmon-tr-line-coupling}, and comment on how the method can be extended to other superconducting qubits. To guide the derivation, we follow a Hamiltonian mechanics approach (for an introduction to Hamiltonian mechanics in the context of electromagnetics, we refer readers to \cite{chew2021qme-made-simple,chew2016quantum,chew2016quantum2}). Since a Hamiltonian mechanics analysis of a transmission line is not as commonly encountered, we begin in Section \ref{subsec:hamiltonian-transmission-line} by reviewing the basic process for an isolated transmission line to introduce physical quantities that will be used throughout this work. Following this, we show in Section \ref{subsec:hamiltonian-transmon} how a similar Hamiltonian analysis approach can be used to arrive at a suitable Schr\"{o}dinger equation to describe a superconducting qubit. Finally, we present the derivation of the self-consistent equations of motion for the coupled system in Section \ref{subsec:semiclassical-eom}.

\subsection{Hamiltonian Mechanics of a Transmission Line}
\label{subsec:hamiltonian-transmission-line}
Formulating the Hamiltonian of an isolated transmission line in terms of voltages and currents is straightforward. The Hamiltonian corresponds to the total energy, which is
\begin{align}
	H_\mathrm{TR} = \frac{1}{2} \int \big[ C \big(V(z,t) \big)^2  + L \big( I(z,t) \big)^2  \big]dz,
	\label{eq:isolated-tr-hamiltonian}
\end{align}
where $V$ is the voltage, $I$ is the current, and $L$ and $C$ are the per-unit-length inductance and capacitance. Unfortunately, much like how electric and magnetic fields are not suitable conjugate variables for a Hamiltonian analysis \cite{chew2021qme-made-simple}, neither are $V$ and $I$. Correspondingly, we must use transmission line parameters more ``like'' the electromagnetic potentials, which are suitable conjugate variables. 

The convention most useful here is to use the node flux $\phi$ and node charge $Q$ \cite{girvin2011circuit,hecht2021engineer}. Physically, the node flux is 
\begin{align}
	\phi(z,t) = \int_{-\infty}^t V(z,\tau) d\tau.
	\label{eq:node-flux-def}
\end{align}
The node flux may also be related to the current as
\begin{align}
	I(z,t) = -L^{-1} \partial_z \phi(z,t).
	\label{eq:i-in-phi}
\end{align}
The node charge $Q$ is the variable that is canonically conjugate to $\phi$. For an isolated transmission line, we have that
\begin{align}
	Q(z,t) = C \partial_t \phi(z,t).
\end{align}
From this and (\ref{eq:node-flux-def}), it is tempting to relate the node charge directly to the voltage as $Q = CV$. Although this is the case for this simple example, it is important to stress that physically the voltage is defined through (\ref{eq:node-flux-def}) in the Hamiltonian formalism, and so the direct relationship of $Q = CV$ will not always hold. This is a consequence of $Q$ being the canonical conjugate variable in the Hamiltonian formalism, which forces it to take on whatever characteristics are needed such that $\phi$ and $Q$ vary in time in tandem to conserve energy \cite{chew2021qme-made-simple}.

Now, in terms of $\phi$ and $Q$, (\ref{eq:isolated-tr-hamiltonian}) becomes 
\begin{align}
	H_\mathrm{TR} = \frac{1}{2}\int \big[ C^{-1} \big( Q(z,t)\big)^2 + L^{-1} \big( \partial_z \phi(z,t) \big)^2    \big] dz.
	\label{eq:isolated-tr-hamiltonian2}
\end{align}
We can now use Hamilton's equations to derive the equations of motion for $\phi$ and $Q$ \cite{chew2021qme-made-simple}, which we will then show to be consistent with the telegrapher's equations \cite{pozar2009microwave}. For $\phi$ and $Q$, Hamilton's equations are
\begin{align}
	\partial_t\phi = \frac{\delta H_\mathrm{TR}}{\delta Q}, \,\,\,\, \partial_t Q = - \frac{\delta H_\mathrm{TR}}{\delta \phi}.
	\label{eq:ham-tr}
\end{align}
Assuming we have boundary conditions that cause the boundary terms that arise from integration by parts to vanish, the requisite functional derivatives can be evaluated to find
\begin{align}
	\partial_t \phi = C^{-1} Q,
	\label{eq:phi-1}
\end{align}
\begin{align}
	\partial_t Q = L^{-1} \partial_z^2\phi.
	\label{eq:q-1}
\end{align}
We can combine (\ref{eq:phi-1}) and (\ref{eq:q-1}) to get
\begin{align}
	\partial_z^2 \phi - LC \partial_t^2\phi = 0,
	\label{eq:phi-wave1}
\end{align}
which we can readily recognize as a wave equation with propagation speed $v = 1/\sqrt{LC}$.

To see that (\ref{eq:phi-wave1}) is consistent with the telegrapher's equations, we begin by recalling that the telegrapher's equations are \cite{pozar2009microwave}
\begin{align}
	\partial_z V = -L \partial_t I,
	\label{eq:kvl}
\end{align} 
\begin{align}
	\partial_z I = -C \partial_t V.
	\label{eq:kcl}
\end{align}
Starting with (\ref{eq:kcl}), we can use (\ref{eq:node-flux-def}) and (\ref{eq:i-in-phi}) to rewrite this as
\begin{align}
	-L^{-1}\partial_z^2\phi = -C\partial_t^2 \phi.
\end{align}
This can be easily rearranged to be seen to be equivalent with (\ref{eq:phi-wave1}), as expected. Substituting (\ref{eq:node-flux-def}) and (\ref{eq:i-in-phi}) in (\ref{eq:kvl}), we find 
\begin{align}
	\partial_z\partial_t \phi = \partial_t\partial_z\phi,
\end{align}
which will always hold for practical cases of interest. Hence, solving the wave equation (\ref{eq:phi-wave1}) is consistent with traditional transmission line theory. Although somewhat circular, this consistency check is the standard way to justify if a Hamiltonian derivation is correct \cite{chew2021qme-made-simple}. The power of the Hamiltonian approach comes in extending a theory to more complicated cases where a first-principles description may not already exist, as will be done in Section \ref{subsec:semiclassical-eom}.

\subsection{Hamiltonian Mechanics of Superconducting Qubits}
\label{subsec:hamiltonian-transmon}
In this section, we will treat the case of a transmon qubit in detail before briefly discussing the extension of the approach to a fluxonium qubit. Traditionally, the Hamiltonian operator for a transmon is most commonly written as
\begin{align}
	{H}_\mathrm{S} = 4E_C\big(\hat{n} - n_g \big)^2 - E_J \cos\hat{\varphi},
	\label{eq:transmon-hamiltonian}
\end{align}
where $\hat{n}$ and $\hat{\varphi}$ are the charge and Josephson phase operators of the qubit \cite{koch2007charge,roth2022transmon}. The classical real-valued quantity $n_g$ is the offset charge that can be used to describe how certain voltages not due to the transmission line modify the equilibrium value of charges in the qubit (e.g., from a DC bias or noise source). Finally, $E_C$ and $E_J$ are the charging and Josephson energies of the qubit, respectively. Explicitly, we have that $E_C = e^2/2C_\Sigma$, where $e$ is the electron charge and $C_\Sigma = C_g + C_q$ is the total capacitance to ground from the terminals of the Josephson junction in the qubit (c.f. Fig. \ref{fig:transmon-tr-line-coupling} for capacitance definitions) \cite{koch2007charge,roth2022transmon}. The Josephson energy is related to design parameters of the Josephson junction, which will not need to be considered explicitly in this work. 

The transmon qubit is designed to operate with $E_J/E_C \gg 1$ to be less sensitive to $n_g$, since common sources of noise can be considered as fluctuations in this value \cite{koch2007charge,roth2022transmon}. Due to the reduced sensitivity to $n_g$, this parameter is often neglected in the analysis of transmon qubits. However, for practical ratios of $E_J/E_C$, higher energy levels of the transmon qubit can still be impacted by the value of $n_g$. Hence, we retain this parameter in our model so that the influence of these higher energy levels can be accounted for within the numerical method, although exhaustively exploring the effects of $n_g$ on the system dynamics are outside of the scope of this work.

Now, working with (\ref{eq:transmon-hamiltonian}) directly is not suitable for developing a Maxwell-Schr\"{o}dinger model as these are typically formulated with wavefunctions in a ``coordinate space'' basis \cite{takeuchi2015maxwell,ryu2016finite,xiang2017high,chen2017unified,xiang2019quantum}. To recast our expressions into a manner similar to this, we use the ``phase basis'' of superconducting systems where the phase operator $\hat{\varphi}$ becomes a regular position variable $\varphi$ and the charge operator $\hat{n}$ becomes $-i\partial_\varphi$ \cite{tinkham2004introduction}. Correspondingly, the Hamiltonian operator becomes a differential operator that can be incorporated into a Schr\"{o}dinger equation to describe the qubit system as an effective particle in a potential well \cite{koch2007charge}. In the case of (\ref{eq:transmon-hamiltonian}), the time-dependent Schr\"{o}dinger equation would be
\begin{align}
	\big[4 E_C \big(\!-\!i\partial_\varphi \!- \!n_g \big)^2  - E_J \cos\varphi \big] \psi(\varphi,t) = i\hbar\partial_t \psi(\varphi,t),
	\label{eq:phase-space-hamiltonian1}
\end{align}  
where $\psi$ is a complex-valued wavefunction with a standard probabilistic interpretation. For the cosine potential energy of the transmon, the wavefunction satisfies periodic boundary conditions of $\psi(\varphi,t) = \psi(\varphi+2\pi,t)$ \cite{koch2007charge}.

In order to follow a Hamiltonian mechanics derivation of the coupled semiclassical equations of motion, we need to determine a Hamiltonian that will yield (\ref{eq:phase-space-hamiltonian1}) through Hamilton's equations. This process is often done in quantum field theory where it is useful to perform a second quantization of a Schr\"{o}dinger wave field \cite{haken1976quantum,cohen1997photons}. Following this example here, albeit without needing to perform a second quantization, the relevant Schr\"{o}dinger wave field Hamiltonian is
\begin{multline}
	H_\mathrm{S} = \int_{-\pi}^{\pi} \frac{1}{i\hbar} \Pi(\varphi,t) \bigg[ 4 E_C \big(\!-\!i\partial_\varphi  \!- \!n_g \big)^2 \\  - E_J \cos\varphi \bigg] \psi(\varphi,t) d\varphi.
	\label{eq:schro-wave1}
\end{multline}
In (\ref{eq:schro-wave1}), we have that $\Pi = i\hbar \psi^*$ is the conjugate function to $\psi$ \cite{haken1976quantum,cohen1997photons}. Further, Hamilton's equations for $\psi$ and $\Pi$ are
\begin{align}
	\partial_t \psi = \frac{\delta H_\mathrm{S}}{\delta \Pi}, \,\,\,\, \partial_t\Pi = -\frac{\delta H_\mathrm{S}}{\delta \psi}.
	\label{eq:ham-qubit}
\end{align}
Evaluating the first functional derivative, we readily find that
\begin{align}
	\partial_t \psi(\varphi,t) = \frac{1}{i\hbar}\big[4 E_C \big(\!-\!i\partial_\varphi \!- \!n_g \big)^2  \!- \!E_J \cos\varphi \big] \psi(\varphi,t),
	\label{eq:phase-space-hamiltonian2}
\end{align}
which is equivalent to (\ref{eq:phase-space-hamiltonian1}). The equation for $\Pi$ is just the complex conjugate of (\ref{eq:phase-space-hamiltonian2}). Although important for second quantization  \cite{haken1976quantum,cohen1997photons}, this is not needed here.

This approach can be readily extended to consider a fluxonium qubit as well. The circuit description of a fluxonium qubit is similar to the transmon shown in Fig. \ref{fig:transmon-tr-line-coupling}, but with the addition of a large parallel linear inductance $L_q$. Typically, this inductance is formed with an array of Josephson junctions, which has the added benefit of making the operating characteristics of the qubit tunable via an applied magnetic flux \cite{manucharyan2009fluxonium,bao2022fluxonium}. The corresponding fluxonium Hamiltonian is 
\begin{align}
	{H}_\mathrm{S}  = 4 E_C\hat{n} ^2 - E_J \cos\big(\hat{\varphi}+\varphi_\mathrm{ext}\big) + \frac{1}{2}E_L \hat{\varphi}^2,
	\label{eq:free-fluxonium1}
\end{align}
where $E_L = (\hbar/(2 e))^2/L_q$ is the inductive energy of the qubit and $\varphi_\mathrm{ext}$ is related to the external magnetic flux used to tune the qubit operating characteristics \cite{krantz2019quantum,manucharyan2009fluxonium}. 

The Hamiltonian in (\ref{eq:free-fluxonium1}) can be rewritten into the phase basis and incorporated into a Schr\"{o}dinger wave field Hamiltonian like in (\ref{eq:schro-wave1}). Evaluating the equation of motion then yields the time-dependent Schr\"{o}dinger equation of 
\begin{multline}
\partial_t \psi(\varphi,t) = -\frac{1}{i\hbar}\big[4E_C \partial_\varphi^2  + E_J \cos\big(\varphi + \varphi_\mathrm{ext} \big)  \\ -\frac{1}{2}E_L \varphi^2  \big]\psi(\varphi,t) .
\label{eq:schro-eom1}
\end{multline}
Due to the quadratic $\varphi$ dependence in (\ref{eq:schro-eom1}), the domain of $\varphi$ is unbounded and $\psi$ no longer satisfies periodic boundary conditions. For numerical discretization purposes we will truncate this domain to a finite range with an artificial boundary condition, with more details discussed in Section \ref{sec:discretization}. 

\subsection{Self-Consistent Semiclassical Equations of Motion}
\label{subsec:semiclassical-eom}
We can now consider the full case of Fig. \ref{fig:transmon-tr-line-coupling} by generalizing the results of \cite{koch2007charge,roth2021macroscopic}. The Hamiltonian is 
\begin{align}
	H = H_\mathrm{TR} + H_\mathrm{S} +  H_\mathrm{I},
	\label{eq:full-ham}
\end{align}
where $H_\mathrm{TR}$ is given in (\ref{eq:isolated-tr-hamiltonian2}), $H_\mathrm{S}$ is given in (\ref{eq:schro-wave1}), and $H_\mathrm{I}$ is the interaction Hamiltonian. For this kind of capacitive coupling, the interaction Hamiltonian is often expressed as \cite{koch2007charge}
\begin{align}
	H_\mathrm{I} = 2e \beta C^{-1} Q(z_0,t) \hat{n},
	\label{eq:h-int}
\end{align}
where $\beta = C_g/C_\Sigma$ is a voltage divider and the transmon is coupled to the transmission line at $z = z_0$. Translating this into a format compatible with $H_\mathrm{TR}$ and $H_\mathrm{S}$, we have 
\begin{multline}
	H_\mathrm{I} = \int\int_{-\pi}^{\pi} \frac{2e\beta}{i\hbar C} Q(z,t) \delta(z-z_0) \\ \times \Pi(\varphi,t) \big(\!-\!i\partial_\varphi \psi(\varphi,t) \big) d\varphi dz.
	\label{eq:h-int2}
\end{multline}

For completeness, it should be noted that to arrive at the simple $H_\mathrm{I}$ in (\ref{eq:h-int}) that only includes $\beta C^{-1}$ it has been assumed that $C \gg C_q, C_g$. For typical devices, $C$ is $O(100 \, \mathrm{pF/m})$ while $C_q$ and $C_g$ are $O(10 \, \mathrm{fF})$ so that the error in the underlying approximation would correspond to modifying $C$ in (\ref{eq:h-int}) by a relative value of $O(10^{-5})$ or smaller, and so is safe to neglect. The same approximation is also needed to keep $H_\mathrm{TR}$ as given in (\ref{eq:isolated-tr-hamiltonian2}) rather than requiring an inconvenient and negligible local change to $C$ at the location of the coupling. More details on a rigorous Hamiltonian treatment of related issues in the context of quantizing superconducting circuit systems in terms of transmission line mode expansions can be found in \cite{parra2018quantum}. Since we consider classical transmission lines without mode expansions, the expression given in (\ref{eq:h-int2}) is adequate for the current purposes.

The coupled equations of motion for this system can now be found using the Hamilton's equations of (\ref{eq:ham-tr}) and (\ref{eq:ham-qubit}), with the adjustment of $H_\mathrm{TR}$ and $H_\mathrm{S}$ to the full Hamiltonian $H$ given in (\ref{eq:full-ham}). For the transmission line, we find that
\begin{align}
	\partial_t Q = L^{-1}\partial_z^2\phi,
	\label{eq:Q-eom}
\end{align}
\begin{align}
	\partial_t \phi = C^{-1}Q + C^{-1} \delta(z-z_0) 2e\beta \langle {n}(t) \rangle,
	\label{eq:phi-eom}
\end{align}
where the expectation value of the charge operator $\hat{n}$ is
\begin{align}
	\langle {n}(t) \rangle = \int_{-\pi}^\pi \psi^*(\varphi,t) \big(\!-\!i\partial_\varphi \psi(\varphi,t) \big) d\varphi.
\end{align}

Combining (\ref{eq:phi-eom}) and (\ref{eq:Q-eom}), we get a wave equation for $\phi$ of
\begin{align}
	\partial_z^2 \phi - LC\partial_t^2\phi = -\delta(z-z_0) L 2e\beta \partial_t \langle {n}(t)\rangle.
	\label{eq:phi-coupled-wave}
\end{align}
Like other Maxwell-Schr\"{o}dinger models, the right-hand side of (\ref{eq:phi-coupled-wave}) can be interpreted as a semiclassical current source due to changes in the expectation value of qubit charge. This matches the expectation that the coupling from the quantum system into the classical one in a semiclassical method involves an expectation value of a current operator \cite{sindelka2010derivation}. 

Next, the equation of motion for $\psi$ is
\begin{multline}
	\big[4 E_C \big(\!-\!i\partial_\varphi \!- \!n_g \big)^2  - E_J \cos\varphi \\ -i2e\beta C^{-1}Q(z_0,t) \partial_\varphi \big] \psi(\varphi,t)  = i\hbar\partial_t \psi(\varphi,t).
	\label{eq:schro-wave2}
\end{multline}
It is necessary to rewrite the term involving $Q$ in (\ref{eq:schro-wave2}) because we will not have access to this quantity when solving (\ref{eq:phi-coupled-wave}). To address this, we use (\ref{eq:phi-eom}) at $z=z_0$ to rewrite $Q$ in (\ref{eq:schro-wave2}) to get
\begin{multline}
	\big[4 E_C \big(\!-\!i\partial_\varphi \!- \!n_g \big)^2  - E_J \cos\varphi \big] \psi(\varphi,t) 
	-i\big[ 2e\beta \partial_t \phi(z_0,t) \\ - (2e\beta)^2 C^{-1} \langle n(t) \rangle \big]\partial_\varphi \psi(\varphi,t)  = i\hbar\partial_t \psi(\varphi,t).
	\label{eq:schro-wave3}
\end{multline}
The term involving $\langle n(t) \rangle$ leads to a nonlinearity in the Schr\"{o}dinger equation that is inconvenient to deal with numerically. However, for the operating characteristics of most systems, this nonlinear correction is typically significantly smaller than the $\partial_t \phi$ term so that we choose to neglect it here.

More explicitly, typically $C$ will be $O(100 \, \mathrm{pF/m})$ and $\beta$ will be \textapprox0.1 or smaller depending on the purpose of the transmission line (e.g., for controlling vs. measuring the state of the qubit). The transmission line voltage (given by $\partial_t\phi(z_0,t)$) will typically be in the nanovolt to microvolt range, while $\langle n(t) \rangle$ will have a maximum of $O(1)$. Hence, the term with $\langle n(t) \rangle$ will be orders of magnitude smaller than the term involving $\partial_t \phi$. Note that this approximation only neglects how the instantaneous correction of the transmission line voltage due to changes in the qubit charge state affects the qubit charge state. The overall dynamical equations still incorporate the effect of the qubit charge state on the transmission line voltage through the semiclassical current source, which then get fed back to the qubit after advancing the system in time in a leapfrog time marching approach.

Considering this, we drop the $\langle n(t) \rangle$ term in (\ref{eq:schro-wave3}) to get
\begin{multline}
	\big[4 E_C \big(\!-\!i\partial_\varphi \!- \!n_g \big)^2  - E_J \cos\varphi \big] \psi(\varphi,t) 
	\\ -i 2e\beta \big( \partial_t \phi(z_0,t) \big) \partial_\varphi \psi(\varphi,t)  = i\hbar\partial_t \psi(\varphi,t).
	\label{eq:schro-wave4}
\end{multline}
This equation can be solved in tandem with (\ref{eq:phi-coupled-wave}) to describe the semiclassical interactions between the transmon qubit and a transmission line. In the case of a fluxonium qubit, the same interaction Hamiltonian (\ref{eq:h-int}) applies so the process of this section can be easily repeated. The resulting time-dependent Schr\"{o}dinger equation to be solved becomes
\begin{multline}
-\bigg[4E_C \partial_\varphi^2  + E_J \cos\big(\varphi + \varphi_\mathrm{ext} \big)  -\frac{1}{2}E_L \varphi^2 \bigg] \psi(\varphi,t) \\ -i 2e\beta \big( \partial_t \phi(z_0,t) \big) \partial_\varphi \psi(\varphi,t)  = i\hbar\partial_t \psi(\varphi,t).
\label{eq:schro-flux}
\end{multline}
\section{Discretization}
\label{sec:discretization}
In this section, we discuss the development of a Maxwell-Schr\"{o}dinger numerical method to solve (\ref{eq:phi-coupled-wave}) and (\ref{eq:schro-wave4}) together. In particular, we utilize a first-order finite element time domain (FETD) method to discretize both sets of equations and advance them in time with leapfrog time marching. Given the one-dimensional spatial nature of the two equations, a finite element discretization is excessive from an accuracy perspective. We use this approach here so that the matrix system has a similar structure to what would be encountered when adapting this method to a full-wave Maxwell-Schr\"{o}dinger method, where the advantages of FETD are clear.

The remainder of this section is organized as follows. In Section \ref{subsec:fetd-full}, we describe the FETD discretization and time marching procedure for solving (\ref{eq:phi-coupled-wave}) and (\ref{eq:schro-wave4}) together. We also briefly discuss the changes needed to consider the fluxonium qubit. Following this, we discuss in Section \ref{subsec:reduced-eigenmode} a strategy for improving the efficiency by discretizing the time-dependent Schr\"{o}dinger equation in terms of eigenstates of the free Hamiltonian operator. This approach can also be used for the fluxonium qubit, and is of particular interest for extending these methods to handling multiple qubits simultaneously, which will be considered in future work. Finally, we present a stability analysis in Section \ref{subsec:stability} of the time marching approaches for the two discretization strategies discussed.

\subsection{Finite Element Time Domain Discretization}
\label{subsec:fetd-full}
To begin, we will consider the FETD discretization of (\ref{eq:phi-coupled-wave}). This follows a standard FETD process (e.g., see \cite{jin2015finite}), however, the boundary conditions for $\phi$ needed in this work require a brief discussion for completeness. For the case of a terminating resistance $R_L$ at the end of a transmission line, Ohm's law in terms of the node flux is
\begin{align}
	\partial_z \phi(z_L,t) = -\frac{L}{R_L} \partial_t \phi(z_L,t),
\end{align}  
where $z_L$ is the $z$-coordinate of $R_L$. To excite the transmission line, a voltage source $V_S(t)$ in series with a source resistance $R_S$ can be re-expressed through a Norton equivalent. In this case, Kirchoff's current law becomes
\begin{align}
	\partial_z\phi(z_S,t) = -\frac{L}{R_S} \big[ V_S(t) - \partial_t \phi(z_S,t)  \big],
\end{align}
where $z_S$ is the $z$-coordinate of the source.

We can now perform the spatial discretization of (\ref{eq:phi-coupled-wave}) by representing $\phi$ as
\begin{align}
	\phi(z,t) = \sum_{n=1}^{N_\phi} \phi_n(t) N_n(z) ,
\end{align}
where $N_n$ is the standard first-order nodal basis function (i.e., a triangular function \cite{jin2015finite}) and $\phi_n$ is the time-dependent expansion coefficient. The weak form of (\ref{eq:phi-coupled-wave}) can be easily found by testing the equation with the same spatial functions. The resulting semi-discrete equation is
\begin{align}
	[T] \frac{d^2}{dt^2} \{\phi\} + [R] \frac{d}{dt}\{\phi\} + [S]\{\phi\} = \{f\},
	\label{eq:phi-semi-discrete}
\end{align}
where $\{\phi\} = [\phi_1, \, \phi_2, \, \ldots , \phi_{N_\phi} ]^\mathrm{T}$. Further,  
\begin{align}
	[T]_{mn} = \int N_m(z) N_n(z) dz,
\end{align}
\begin{align}
	[R]_{mn} = \frac{L}{R_L} N_m(z_L) N_n(z_L) + \frac{L}{R_S} N_m(z_S) N_n(z_S),
\end{align}
\begin{align}
	[S]_{mn} = \int \big( \partial_z N_m(z)\big) \big( \partial_z N_n(z) \big) dz,
\end{align}
\begin{align}
	\{f\}_m =   \frac{L}{R_S} N_m(z_S) V_s(t) + N_m(z_0) L 2e \beta \partial_t \langle n(t) \rangle. 
	\label{eq:phi-exc}
\end{align}
We will discuss the temporal discretization of (\ref{eq:phi-semi-discrete}) after discussing the spatial discretization of (\ref{eq:schro-wave4}).

The spatial discretization of (\ref{eq:schro-wave4}) can follow a similar pattern. To begin, we first expand all the terms out to have 
\begin{multline}
	\big[ \!-\!4E_C \partial_\varphi^2 +i8 n_g E_c \partial_\varphi  + 4 E_c n_g^2  - E_J \cos\varphi \big] \psi(\varphi,t) \\
	-i2e\beta \big( \partial_t \phi(z_0,t)\big) \partial_\varphi \psi(\varphi,t) = i\hbar \partial_t \psi(\varphi,t).
	\label{eq:transmon-full}
\end{multline}
Expanding $\psi$ as
\begin{align}
	\psi(\varphi,t) = \sum_{n=1}^{N_\psi}   a_n(t) N_n(\varphi)
\end{align}
and testing (\ref{eq:transmon-full}) with the same spatial functions, the semi-discrete version of (\ref{eq:transmon-full}) becomes
\begin{align}
	[E] \frac{d}{dt} \{a\} = \frac{1}{i\hbar} [H_0] \{a\} - \frac{2e\beta}{\hbar}  \big( \partial_t \phi(z_0,t)  \big) [Q] \{a\} ,
	\label{eq:schro-semi-discrete}
\end{align}
where $\{a\} = [a_1, \, a_2, \, \ldots , a_{N_\psi} ]^\mathrm{T}$ and
\begin{align}
	[H_0] = 4E_C[N] +i8n_g E_C [Q] + 4E_C n_g^2 [E] - E_J [V] ,
\end{align}
\begin{align}
	[E]_{mn} = \int_{-\pi}^\pi N_m(\varphi) N_n(\varphi) d\varphi,
\end{align}
\begin{align}
	[N]_{mn} = \int_{-\pi}^\pi \big( \partial_\varphi N_m(\varphi)  \big) \big( \partial_\varphi N_n(\varphi)   \big) d\varphi,
\end{align}
\begin{align}
	[Q]_{mn} = \int_{-\pi}^\pi N_m(\varphi) \big( \partial_\varphi N_n(\varphi)   \big) d\varphi,
\end{align}
\begin{align}
	[V]_{mn} = \int_{-\pi}^\pi N_m(\varphi) N_n(\varphi) \cos(\varphi) d\varphi.
\end{align}

The spatial discretization of (\ref{eq:schro-flux}) can be handled similarly, where we omit the explicit expressions for brevity. The main distinction is that the fluxonium potential energy is not periodic. To truncate the discretization region, we apply a homogeneous Dirichlet boundary condition at both sides that corresponds physically to the potential energy ``jumping'' to an infinite value. Due to the fast growth in the quadratic part of the fluxonium potential well, this boundary condition has little effect on the wavefunction if the discretization interval is wide enough. Here, we discretize over the interval $\varphi \in [-6\pi, 6\pi]$ to ensure the wavefunctions of interest go to 0 well before reaching the artificial terminating boundary condition.

Now, to solve (\ref{eq:phi-semi-discrete}) and (\ref{eq:schro-semi-discrete}) together we use leapfrog time marching. To achieve this, we represent $\phi$ and $\psi$ on staggered temporal grids that are offset by a half time step. Choosing $\phi$ to be represented at integer time steps, we can discretize the temporal derivatives in (\ref{eq:phi-semi-discrete}) using standard central difference formulas \cite{jin2015finite}. The resulting time stepping equation is then
\begin{multline}
	\bigg( \frac{[T]}{(\Delta t)^2} + \frac{[R]}{2\Delta t} \bigg) \{\phi^{(j+1)}\} = \bigg( \frac{2[T]}{(\Delta t)^2} - [S] \bigg) \{\phi^{(j)}\} \\ -\bigg( \frac{[T]}{(\Delta t)^2} - \frac{[R]}{2\Delta t} \bigg)\{\phi^{(j-1)}\} + \{f^{(j)}\},
	\label{eq:phi-time-stepping}
\end{multline}
where $\Delta t$ is the time step and superscript $(j)$ denotes a vector of coefficients evaluated at $t = j\Delta t$ with $j$ an integer. Note that in evaluating the $\partial_t \langle n(t) \rangle$ term in (\ref{eq:phi-exc}), we use a central difference formula that uses a time step size of $\Delta t / 2$ evaluated at $t = j\Delta t$. Explicitly, we have that
\begin{align}
	\partial_t \langle n^{(j)} \rangle = \frac{\langle n^{(j+1/2)} \rangle - \langle n^{(j-1/2)} \rangle }{\Delta t},
\end{align}
which samples $\langle n(t) \rangle$ at known temporal values of $\psi$.

A similar process also works for the temporal discretization of (\ref{eq:schro-semi-discrete}), but with $\psi$ sampled on the staggered temporal grid. Using a simple central difference discretization of (\ref{eq:schro-semi-discrete}) at time step $(j+1/2)\Delta t$, we get
\begin{multline}
	[E]\{a^{(j+3/2)}\} = [E] \{a^{(j-1/2)}\} + \frac{2\Delta t}{i\hbar} [H_0]\{a^{(j+1/2)}\} \\
	-\frac{4e\beta\Delta t }{\hbar} \big( \partial_t \phi^{(j+1/2)}(z_0)  \big) [Q] \{a^{(j+1/2)}\}.
	\label{eq:schro-full-stepping}
\end{multline}
To evaluate $\partial_t \phi^{(j+1/2)}(z_0)$, we use a time step size of $\Delta t/2$ so that it becomes
\begin{align}
	\partial_t \phi^{(j+1/2)}(z_0) = \frac{\phi^{(n+1)}(z_0)-\phi^{(n)}(z_0)}{\Delta t},
\end{align}
which samples $\phi(z_0,t)$ at known temporal values of $\phi$.

\subsection{Reduced Eigenstate Discretization}
\label{subsec:reduced-eigenmode}
In most situations, it is advantageous to consider the dynamics of the qubit in terms of its eigenstates rather than using a full spatial description like that used in Section \ref{subsec:fetd-full}. This can be easily incorporated into a Maxwell-Schr\"{o}dinger discretization, and leads to a more efficient method that also produces more intuitive results. 

To do this, we first must find the eigenstates and eigenenergies of the free qubit Hamiltonian operator; i.e., (\ref{eq:transmon-hamiltonian}) for a transmon and (\ref{eq:free-fluxonium1}) for a fluxonium. These can be found numerically very easily using the finite element method described in Section \ref{subsec:fetd-full}. In particular, the $n$th numerical eigenstate $\{\psi_n\}$ with corresponding eigenenergy $E_n$ can be found as solutions to the generalized eigenvalue problem
\begin{align}
	[H_0] \{\psi_n\} = E_n [E]\{\psi_n\}.
\end{align}
Using these eigenstates, we can then expand $\psi$ as
\begin{align}
	\{\psi(t)\}  = \sum_{n=0}^{N_\mathrm{eig}-1} c_n(t) \{\psi_n\}.
\end{align}
In many situations, it is common to only consider $N_\mathrm{eig} = 3$ for a transmon qubit \cite{krantz2019quantum}. However, as will be discussed in Section \ref{sec:results}, this can lead to significantly incorrect results if one is not careful with the design of the incident pulses on the transmission line. This has also been noted as a deficiency in the quantum control literature for transmon qubits \cite{jones2021approximations}. In the case of a fluxonium qubit, the number of needed eigenstates is more challenging to determine \textit{a priori} due to the strong nonlinearity of the qubit \cite{zhu2013circuit}. In this case, typical numerical convergence studies can be used on a case-by-case basis.

Now, the orthonormality of the qubit eigenstates greatly simplifies the discretization of the time-dependent Schr\"{o}dinger equation (\ref{eq:schro-wave4}) or (\ref{eq:schro-flux}). In particular, we have
\begin{align}
	i\hbar \frac{d}{dt}\{c\} = [\mathcal{E}] \{c\} -i2e\beta \big(\partial_t \phi(z_0,t)\big) [\mathcal{Q}] \{c\},
	\label{eq:red1}
\end{align}
where $\{c\} = [c_0, \, c_1, \, \ldots , \, c_{N_\mathrm{eig}-1}]^\mathrm{T}$. We further have that
\begin{align}
	[\mathcal{E}]_{mn} = \delta_{mn} E_n, 
\end{align}
\begin{align}
	[\mathcal{Q}]_{mn} = \{\psi_m\}^\dagger [Q] \{\psi_n\},
\end{align}
where $\delta_{mn}$ is a Kronecker delta and a superscript $\dagger$ denotes a conjugate transpose. Following the same temporal discretization strategy of Section \ref{subsec:fetd-full}, the time stepping equation for (\ref{eq:red1}) is
\begin{multline}
	\{c^{(j+3/2)}\} = \{c^{(j-1/2)}\} + \frac{2\Delta t}{i\hbar} [\mathcal{E}]\{c^{(j+1/2)}\} \\ -\frac{4e\beta\Delta t}{\hbar} \big( \partial_t \phi^{(j+1/2)}(z_0)\big)[\mathcal{Q}]\{c^{(j+1/2)}\}.
	\label{eq:schro-red-stepping}
\end{multline}
This equation can be solved along with (\ref{eq:phi-time-stepping}) to march the overall system of equations forward in time.

In quantum mechanics, it is typically only the relative energy difference between eigenstates that is important because the energy reference value can be adjusted arbitrarily. In this work, we reference all energies relative to the $E_0$ value, which helps with extracting oscillation frequencies of the different eigenstates in Section \ref{sec:results}.

\subsection{Stability Analysis}
\label{subsec:stability}
As with any time domain method, a stability analysis is necessary to determine what values of $\Delta t$ the system of equations can be safely solved for. The discretization strategies used in Sections \ref{subsec:fetd-full} and \ref{subsec:reduced-eigenmode} lead to conditionally stable systems, for which stability conditions can be derived using standard methods.

To begin, we will consider the wave equation for the node flux, whose time stepping equation was given in (\ref{eq:phi-time-stepping}). This time stepping equation exactly matches the format of a typical full-wave FETD system, and so the same stability condition holds \cite{jiao2002general,jin2015finite}. From a Z-transform analysis, the stability condition is found to be
\begin{align}
	\Delta t \leq \frac{2}{\sqrt{\rho\big([T]^{-1}[S]\big)}},
	\label{eq:phi-stability}
\end{align}
where $\rho\big([A]\big)$ denotes the spectral radius of $[A]$.

The analysis of the stability condition for (\ref{eq:schro-full-stepping}) and (\ref{eq:schro-red-stepping}) can be found following the basic procedure of \cite{soriano2004analysis}. In this approach, a constraint on the temporal eigenvalues of the time-dependent Schr\"{o}dinger equation are found. The temporal eigenvalues are given by
\begin{align}
	i\hbar \partial_t \psi = \lambda \psi,
\end{align}
where we use a generic notation of $\psi$ since the approach works the same for (\ref{eq:schro-full-stepping}) and (\ref{eq:schro-red-stepping}). After applying the central difference formula, we have
\begin{align}
	i\hbar \frac{\psi^{(j+1)}-\psi^{(j-1)}}{2\Delta t} = \lambda \psi^{(n)}.
	\label{eq:stability1}
\end{align}
An amplification factor can be defined as $g = \psi^{(n+1)}/\psi^{(n)} = \psi^{(n)}/\psi^{(n-1)} $, where stability will only occur if $|g|\leq 1$. The amplification factor can be substituted into (\ref{eq:stability1}) and $g$ can be solved for as
\begin{align}
	g = -i \frac{\lambda \Delta t}{\hbar} \pm \sqrt{1-\bigg(  \frac{\lambda \Delta t}{\hbar} \bigg)^2}.
\end{align}
For $|g|\leq 1$, we find that our stability condition is
\begin{align}
	\Delta t \leq \frac{\hbar}{\lambda}.
\end{align}

Now, the manner of computing $\lambda$ depends on whether (\ref{eq:schro-full-stepping}) or (\ref{eq:schro-red-stepping}) is being solved. For (\ref{eq:schro-full-stepping}), $\lambda$ is related to the eigenvalues of the matrix system on the right-hand side of (\ref{eq:schro-semi-discrete}) after multiplying by $[E]^{-1}$. Since this matrix system changes depending on the value of $\partial_t\phi(z_0,t)$, it is not possible to find a ``true'' stability condition that is always valid. Instead, we make an estimate for the stability condition by using the maximum value expected for $\partial_t \phi(z_0,t)$ to occur in a given simulation, which can be readily-inferred from the temporal profile of the voltage source or by solving the transmission line subsystem in the absence of any qubits. If we denote this maximum value of $\partial_t \phi(z_0,t)$ as $V_\mathrm{max}(z_0)$, then we can estimate the stability condition as
\begin{align}
	\Delta t \leq \frac{\hbar}{\rho \bigg( [E]^{-1} \big( [H_0] -i2e\beta V_\mathrm{max}(z_0) [Q]  \big) \bigg) }
	\label{eq:schro-full-stability}
\end{align}
Similarly, the stability condition for (\ref{eq:schro-red-stepping}) is related to the eigenvalues of the matrix system on the right-hand side of (\ref{eq:red1}). We can estimate the stability condition for (\ref{eq:schro-red-stepping}) as
\begin{align}
	\Delta t \leq \frac{\hbar}{\rho\big( [\mathcal{E}] - i2e\beta V_\mathrm{max}(z_0) [\mathcal{Q}] \big)}.
	\label{eq:schro-red-stability}
\end{align}
We have verified these stability conditions with numerical experiments, but do not show details of this for brevity.

Although the forms of the stability conditions (\ref{eq:schro-full-stability}) and (\ref{eq:schro-red-stability}) are very similar, in practice they produce very different numerical values. Generally, we find that (\ref{eq:schro-red-stability}) leads to a stability condition that is around $O(10^3)$ larger than that of (\ref{eq:schro-full-stability}). Given that (\ref{eq:schro-full-stability}) is also often $O(10^2)$ smaller than (\ref{eq:phi-stability}), this represents a significant advantage of using the reduced eigenstate expansion approach for forming the Maxwell-Schr\"{o}dinger system of equations. It is expected that further improvements in efficiency can occur in the future by exploring more sophisticated time stepping algorithms with higher-order accuracy or different stability constraints (e.g., Newmark-beta or Crank-Nicolson schemes).
\section{Numerical Results}
\label{sec:results}
In this section, we present numerical results to qualitatively and quantitatively validate the accuracy of the proposed formulation. To facilitate this, we consider simple systems for which various kinds of theoretical analysis are amenable. We begin in Section \ref{subsec:qubit-control} by considering the control of superconducting qubits with classical microwave drives, where we also investigate the comparison between the discretization techniques discussed in Section \ref{sec:discretization}. Following this, we provide quantitative validation of the proposed Maxwell-Schr\"{o}dinger method by modeling dispersive regime effects relevant to qubit state measurement in Section \ref{subsec:qubit-measurement}. In each section, we review the basic details of the theoretical predictions that we use in validating our numerical method prior to discussing the numerical results. 

In each simulation, the following parameters were used. The transmission lines always have $L = 0.7125 \, \mu\mathrm{H/m}$ and $C = 285 \, \mathrm{pF/m}$, which corresponds to a $50\,\Omega$ impedance. This value for $C$ was estimated from analytical calculations of coplanar waveguides (expressions available, e.g., in \cite{simons2004coplanar}) for common device parameters taken from \cite{goppl2008coplanar,houck2007generating}. Similarly, for all simulations involving a transmon qubit $n_g = 0.5$. 

Finally, a full analysis of computational speed is outside of the scope of this work, but a typical Maxwell-Schr\"{o}dinger simulation was completed in $O(10 \, \mathrm{s})$ on a standard workstation without utilizing any parallelization or significant code optimization. We find these simulation times to be comparable to the standard quantum time evolution methods used in QuTiP \cite{johansson2012qutip} that do not include self-consistent interactions if open quantum system effects are neglected. Attempting to include self-consistent interactions in QuTiP rapidly becomes computationally prohibitive due to the exponential growth of the state space and prevented using this tool for quantitative validation of the proposed Maxwell-Schr\"{o}dinger method. As a result, we have focused on quantitatively validating our method primarily through comparison to amenable theoretical results that are detailed in Section \ref{subsec:qubit-measurement}. 


\subsection{Qubit Control}
\label{subsec:qubit-control}
A natural application of Maxwell-Schr\"{o}dinger methods is to analyze the fidelity of control pulses. Here, we illustrate simple examples of a transmon or fluxonium qubit controlled through driven Rabi oscillations \cite{krantz2019quantum}. In this approach, a modulated Gaussian pulse with center frequency matching the first transition frequency of the qubit is applied. From a simplified theoretical treatment, the final qubit state will vary in a cyclic manner between the ground and first excited state depending on the area of the baseband Gaussian pulse \cite{fox2006quantum}. Sufficiently arbitrary output qubit states for practical applications can be achieved by controlling the pulse area and relative phase of the applied microwave pulses \cite{krantz2019quantum}. Although simple, the stringent control fidelity requirements that can be on the order of 0.9999 \cite{huang2019performance} can make achieving this at faster operating speeds difficult for transmon qubits due to their weak nonlinearity and for fluxonium qubits due to their complicated structure.

\begin{figure}[t!]
	\centering
	\includegraphics[width=\linewidth]{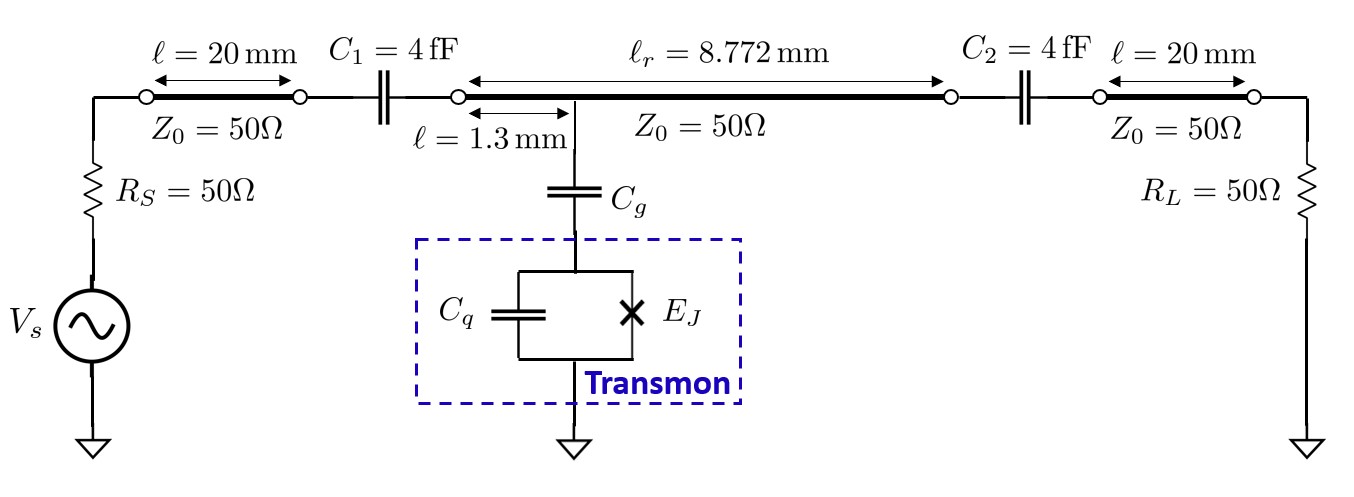}
	\caption{Circuit schematic used for transmon control analysis. Bold lines denote transmission lines, while thin lines denote a regular circuit connection. Parameters that do not have quantities explicitly given are discussed in the main text.}
	\label{fig:device-schematic}
\end{figure}

The first system we will analyze is shown in Fig. \ref{fig:device-schematic}, which consists of a transmon coupled to a half-wavelength resonator. The device parameters are loosely based on a symmetrized version of the single photon source discussed in \cite{houck2007generating}. In all simulations of this device, we use $E_J/E_C = 25$ and maintain $C_\Sigma = C_g+C_q = 55 \, \mathrm{fF}$ to ensure that the transition frequencies of the transmon remain the same regardless of the coupling strength $\beta$ used in a particular simulation. For reference, the first transition frequency of the transmon is $4.60 \, \mathrm{GHz}$, the second transition frequency is $4.14 \, \mathrm{GHz}$, and the first resonance frequency of the transmission line is $4.00 \, \mathrm{GHz}$.

Here, we calibrate a $2\pi$-pulse that transitions the qubit from the ground state to the first excited state and then back to the ground state using numerical experiments and then use this result to extrapolate to other pulse areas. Initially, we use a modulated Gaussian pulse with a standard deviation of $\sigma = 5 \, \mathrm{ns}$ and $\beta = 0.01$. This $\beta$ is small for the topology of Fig. \ref{fig:device-schematic}, but keeps the interaction between the two subsystems minor enough that we can qualitatively validate our results against a simpler treatment that does not consider the interaction self-consistently. In particular, the simpler approach neglects the effect of the semiclassical current source in (\ref{eq:phi-coupled-wave}) to only allow one-way coupling of the transmission line signals onto the transmon. This ensures that all propagation effects are accounted for in delivering the drive to the transmon, but is otherwise equivalent to a typical quantum control analysis that neglects open quantum system effects \cite{jones2021approximations,johansson2012qutip}.

\begin{figure}[t!]
	\centering
	\includegraphics[width=0.95\linewidth]{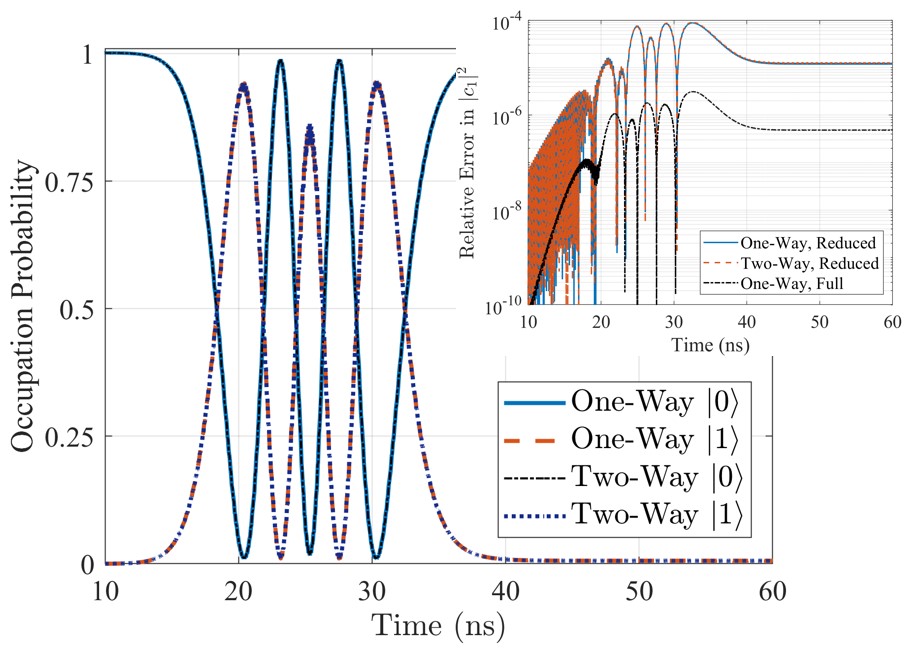}
	\caption{Occupation probabilities of the ground and first excited states (denoted as $|0\rangle$ and $|1\rangle$ in Dirac bra-ket notation) for a $6\pi$-pulse when $\beta = 0.01$ and $\sigma = 5 \, \mathrm{ns}$. Cases considered include when self-consistent interactions are (two-way) or are not (one-way) considered. All results in the main figure use the full FETD discretization (full). The inset shows that the relative error is very small for all methods including the reduced eigenstate discretization (reduced), where the two-way method with full FETD discretization is the reference. To achieve convergence between reduced eigenstate and full FETD discretizations, $N_\mathrm{eig} = 3$ was required.}
	\label{fig:rabi_b0p01_s5_6pi}
\end{figure}

To show our methods match expected theoretical results, we present the numerical results of the various methods discussed in this work for a $6\pi$-pulse in Fig. \ref{fig:rabi_b0p01_s5_6pi}. This pulse area should complete three full Rabi cycles and end with the qubit in the ground state. We see that all of the numerical results demonstrate this behavior and maintain close agreement throughout the entire course of the simulation. For the simulations using the reduced eigenstate discretization of Section \ref{subsec:reduced-eigenmode}, we used $N_\mathrm{eig} = 3$. This is generally considered to be the minimum number of eigenstates needed to describe the dynamics of a transmon, and we see that in this simple scenario this is indeed adequate. However, higher convergence to the full FETD discretization can be achieved by including more eigenstates in the simulation.

\begin{figure}[t!]
	\centering
	\includegraphics[width=0.95\linewidth]{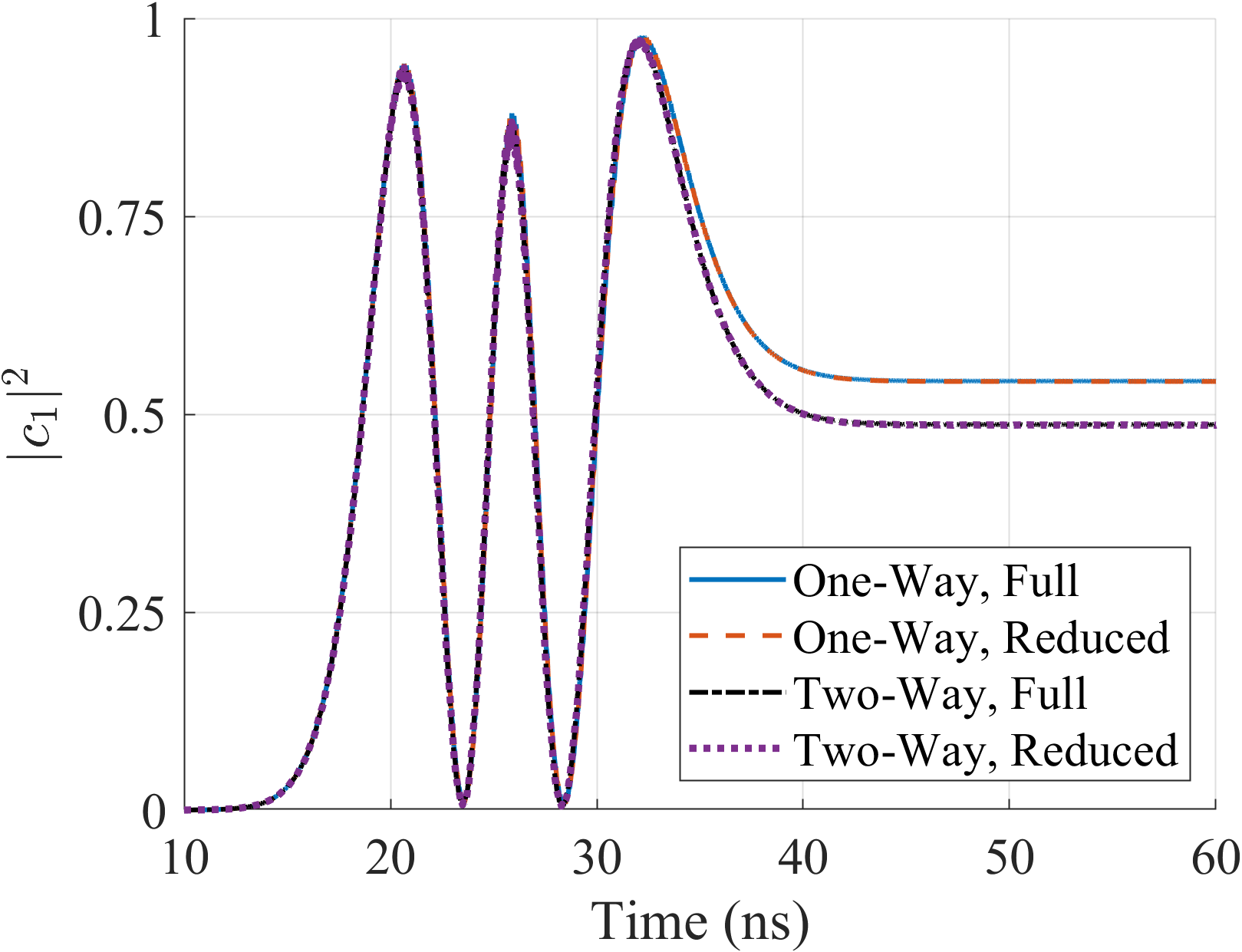}
	\caption{First excited state occupation probability for a $5.5\pi$-pulse when $\beta = 0.1$ and $\sigma = 5 \, \mathrm{ns}$. The increased coupling between the systems results in the deviation between the methods with two-way vs. one-way coupling. To achieve convergence between reduced eigenstate and full FETD discretizations, $N_\mathrm{eig} = 4$ was required.}
	\label{fig:rabi_b0p1_s5_5p5pi}
\end{figure}

Next, we look at less trivial scenarios where the capability of a numerical method is of more value. In particular, we increase $\beta$ to $0.1$, which is a more realistic value for a resonator coupled to a transmon in modern designs (see, e.g., \cite{kosen2022building}). We also keep the Gaussian pulse $\sigma = 5 \, \mathrm{ns}$, but adjust the pulse amplitude to correspond to a $5.5\pi$-pulse that should transition the transmon into an equal superposition of its first two states after 2.5 Rabi cycles. From the results in Fig. \ref{fig:rabi_b0p1_s5_5p5pi}, we see that there begins to be a significant discrepancy in the final state between the self-consistent method and the approach that only allows one-way coupling. This deviation is easily understood to occur due to the back-action of the transmon on the transmission line. Calibrating for these effects is typically handled experimentally, which can become time consuming as system sizes are increased. In these cases, improved numerical methods like a Maxwell-Schr\"{o}dinger model can aid in minimizing the experimental work needed to verify system performance. It should also be noted that for this example $N_\mathrm{eig} = 4$ is necessary to achieve convergence in the dynamics between the various methods. The increased number of eigenstates is also due to the increased coupling between the two subsystems.

\begin{figure}[t!]
	\centering
	\includegraphics[width=0.85\linewidth]{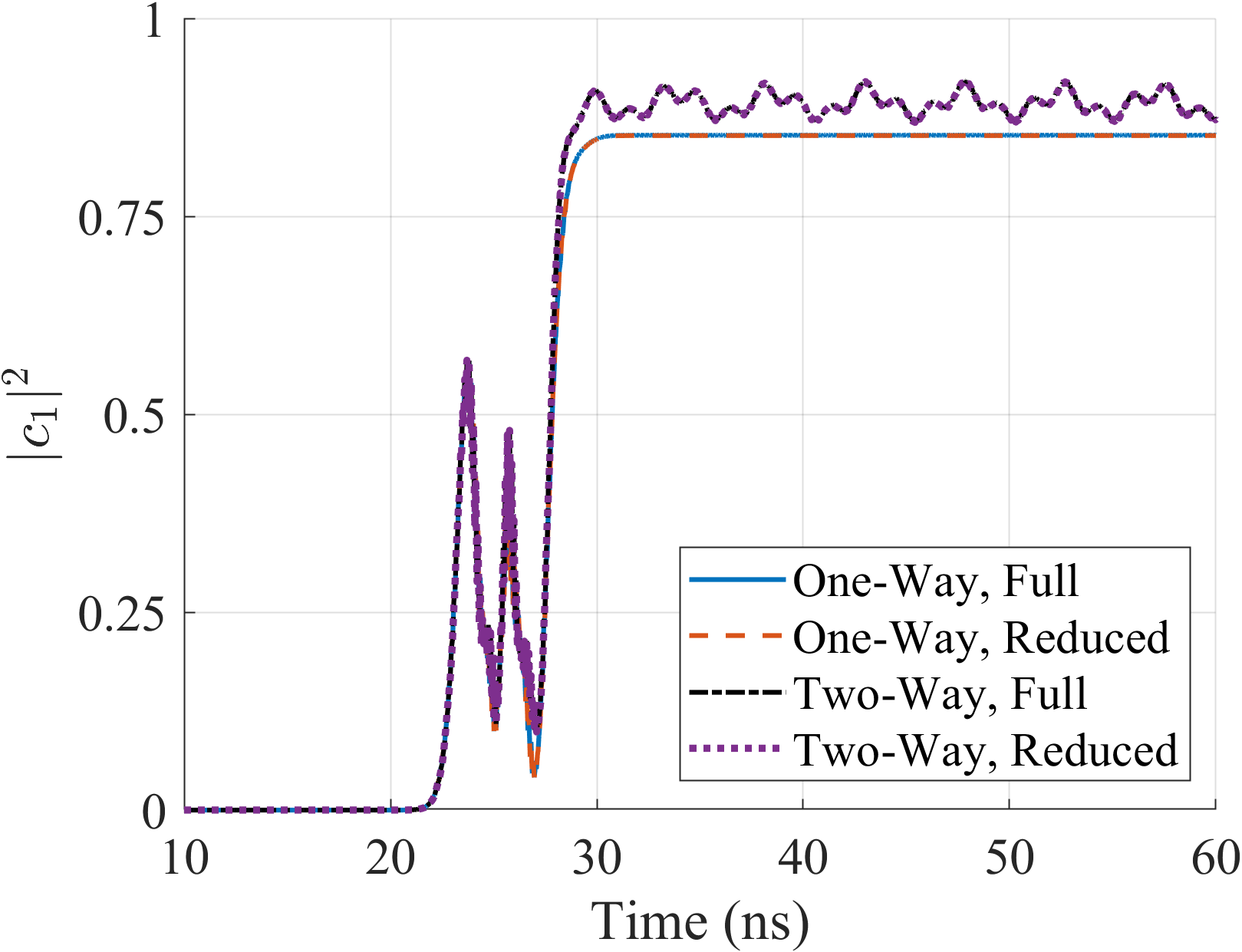}
	\caption{First excited state occupation probability for a $5.5\pi$-pulse when $\beta = 0.1$ and $\sigma = 1.5 \, \mathrm{ns}$. The shorter pulse width results in non-negligible interactions with higher qubit eigenstates and the transmission line, causing significant discrepancies between the methods with two-way vs. one-way coupling. To achieve convergence between reduced eigenstate and full FETD discretizations, $N_\mathrm{eig} = 6$ was required.}
	\label{fig:rabi_b0p1_s1p5_5p5pi}
\end{figure}

Now, if we naively attempt to shorten the duration of the Gaussian pulse to $\sigma = 1.5 \, \mathrm{ns}$ while increasing the pulse amplitude to still achieve a $5.5\pi$-pulse, we see in Fig. \ref{fig:rabi_b0p1_s1p5_5p5pi} that non-trivial dynamical effects begin to emerge. From a simplified theoretical treatment, the final state of the transmon qubit should be identical between Figs. \ref{fig:rabi_b0p1_s5_5p5pi} and \ref{fig:rabi_b0p1_s1p5_5p5pi}. The deviation is due to non-negligible interactions with higher transmon eigenstates, which are neglected in the theoretical treatment. The significant deviations between the results of the self-consistent method and the method that only considers one-way coupling in Fig. \ref{fig:rabi_b0p1_s1p5_5p5pi} is due purely to the back-action of the transmon on the transmission line. The periodicity in these additional dynamics suggests this effect is due to wave interference within the transmission line resonator. Capturing these effects in typical fully-quantum modeling methods is computationally prohibitive due to the many transmission line resonator modes required to capture the wave propagation effects (estimated at \textapprox20). In contrast to this, the computational work for the Maxwell-Schr\"{o}dinger method is unchanged, although to achieve convergence in the dynamics $N_\mathrm{eig} = 6$ was required for this simulation. 

\begin{figure}[t!]
	\centering
	\includegraphics[width=0.85\linewidth]{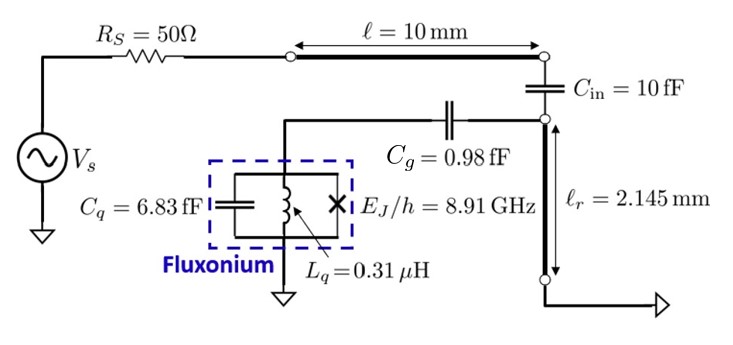}
	\caption{Schematic of the fluxonium system analyzed in this work.}
	\label{fig:fluxonium-device-schematic}
\end{figure}

The need for this many eigenstates in quantum control analysis is not unprecedented, with the results of \cite{jones2021approximations} suggesting that \textapprox10 eigenstates can be needed to achieve convergence in transmon dynamics for realistic pulse designs. An exhaustive study on the number of eigenstates needed for convergence is outside of the scope of this work, but for the modeling of a single qubit the computational cost of including additional eigenstates in the numerical method is negligible. However, as these methods are adapted to consider multi-qubit interactions in the future, a more detailed understanding of the needed number of eigenstates will be valuable to minimize the size of state space the dynamics need to be modeled within.

\begin{figure}[t]
	\centering
	\begin{subfigure}[t]{0.85\linewidth}
		\includegraphics[width=\textwidth]{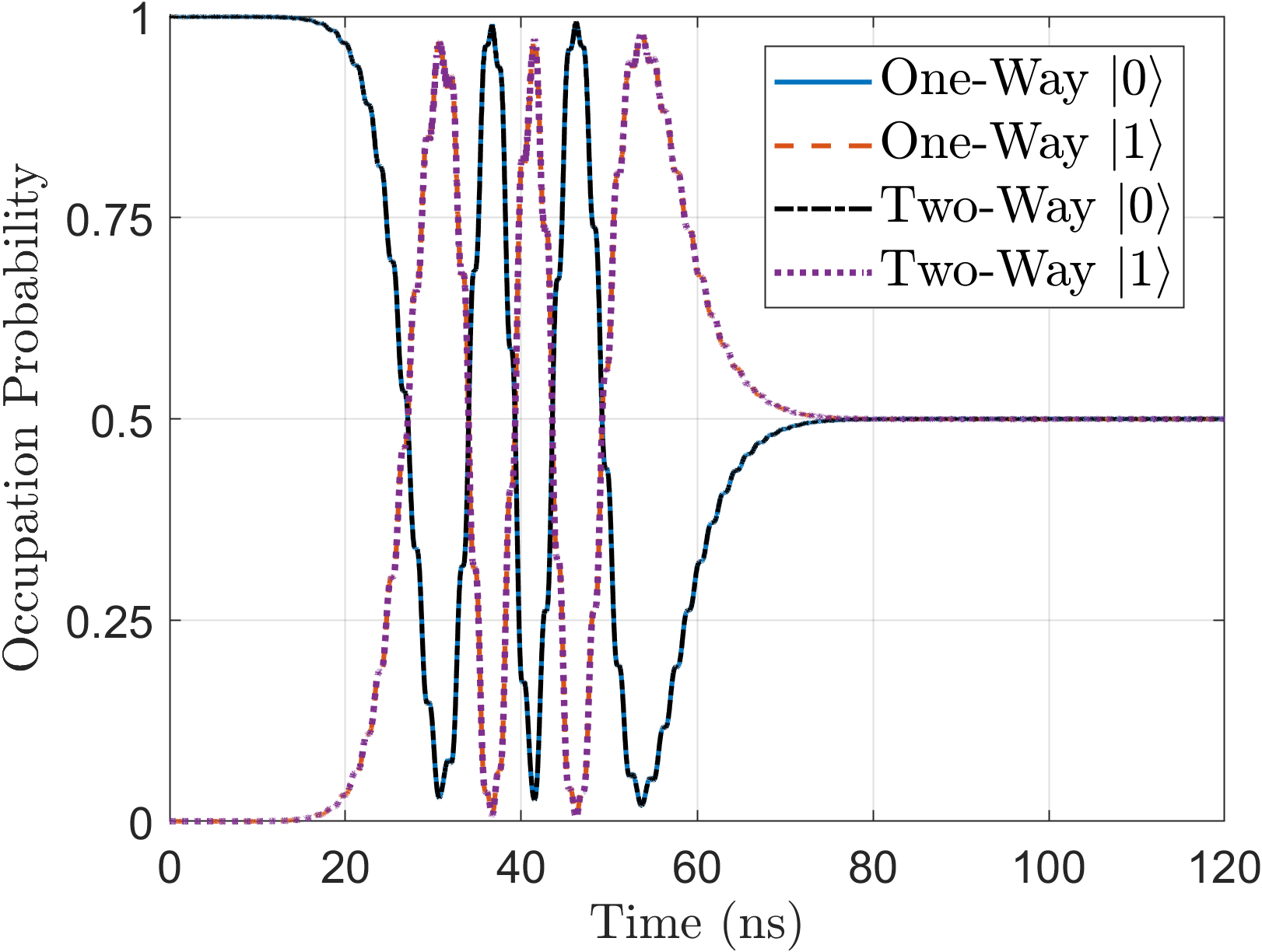}
		\caption{}
		\label{subfig:fluxonium-control-flux0p5}
	\end{subfigure}
	\begin{subfigure}[t]{0.85\linewidth}
		\includegraphics[width=\textwidth]{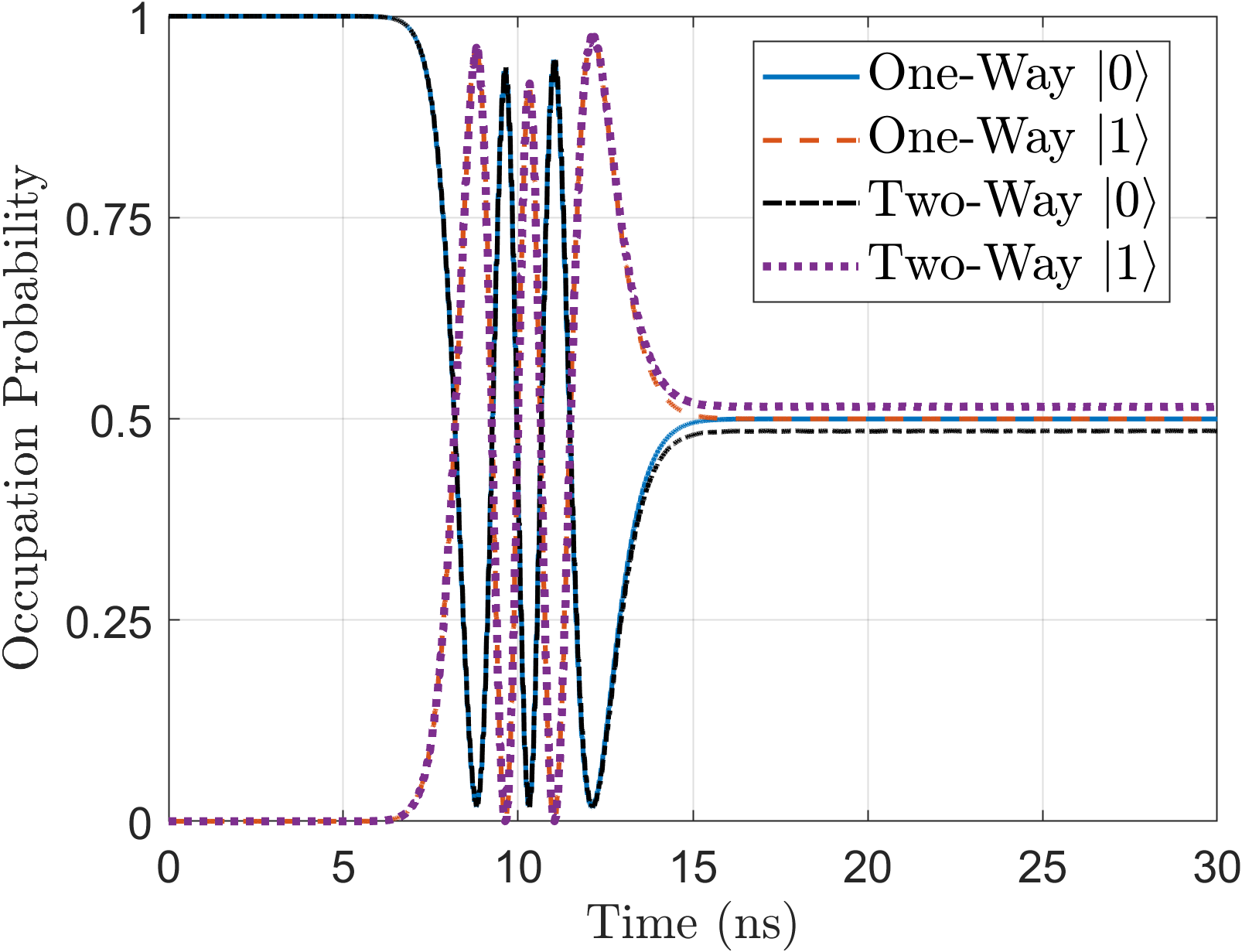}
		\caption{}
		\label{subfig:fluxonium-control-flux0p0}
	\end{subfigure}
	\caption{Occupation probabilities of the fluxonium ground and first excited states for a $5.5\pi$-pulse. The qubit operating points are (a) $\varphi_\mathrm{ext} = -0.5$ and (b) $\varphi_\mathrm{ext} = 0$. At each operating point, the system is modeled with two-way and one-way coupling. For the faster pulse speed in (b), the self-consistent interactions become non-negligible.}
	\label{fig:control-combined}
\end{figure}

Next, we consider the control of a fluxonium qubit for the system shown in Fig. \ref{fig:fluxonium-device-schematic}, which is based on the device analyzed in \cite{manucharyan2009fluxonium,zhu2013circuit}. Here, we consider controlling the qubit using conventional means that do not consider the self-consistent interactions included in a Maxwell-Schr\"{o}dinger model. In particular, we optimize a $5.5\pi$-pulse when only allowing one-way coupling of the signals in the transmission line onto the fluxonium qubit. We then use these same pulse parameters in the full Maxwell-Schr\"{o}dinger model with two-way coupling to observe the impact of the self-consistent interactions. The results of this are shown in Fig. \ref{fig:control-combined} for two different operating points of $\varphi_\mathrm{ext} = -0.5$ and $\varphi_\mathrm{ext}= 0$. For all cases, $N_\mathrm{eig} = 10$ and only the reduced eigenstate discretization was considered.

For the $\varphi_\mathrm{ext} =-0.5$ case, the qubit transition frequency is $368 \, \mathrm{MHz}$, so the pulse speed must be kept relatively slow with $\sigma = 10 \, \mathrm{ns}$ to prevent non-ideal effects occurring due to the pulse bandwidth becoming too comparable to the transition frequency. Due to this slow pulse speed and the significant detuning of the control pulse from the transmission line resonator frequency (first resonance occurs at $8.18 \, \mathrm{GHz}$), the self-consistent interactions included in the Maxwell-Schr\"{o}dinger model are negligible. However, when this procedure is repeated at $\varphi_\mathrm{ext} = 0$, the self-consistent interactions become important. In this case, the qubit transition frequency is $9.17 \, \mathrm{GHz}$, which allows for a significantly faster pulse speed of $\sigma = 1.5 \, \mathrm{ns}$ to be utilized. Although a faster pulse is possible, it is seen that the self-consistent interactions lead to a significant deviation from the intended result considering the control fidelity requirements.

\begin{figure}[t!]
	\centering
	\includegraphics[width=0.9\linewidth]{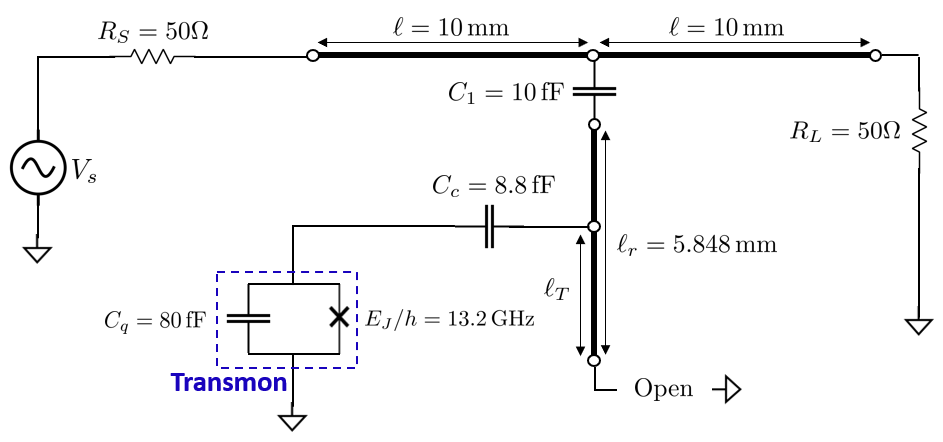}
	\caption{Schematic of the transmon system used to analyze dispersive regime effects in this work.}
	\label{fig:transmon_readout_schematic}
\end{figure}

\begin{figure}[t!]
	\centering
	\includegraphics[width=0.9\linewidth]{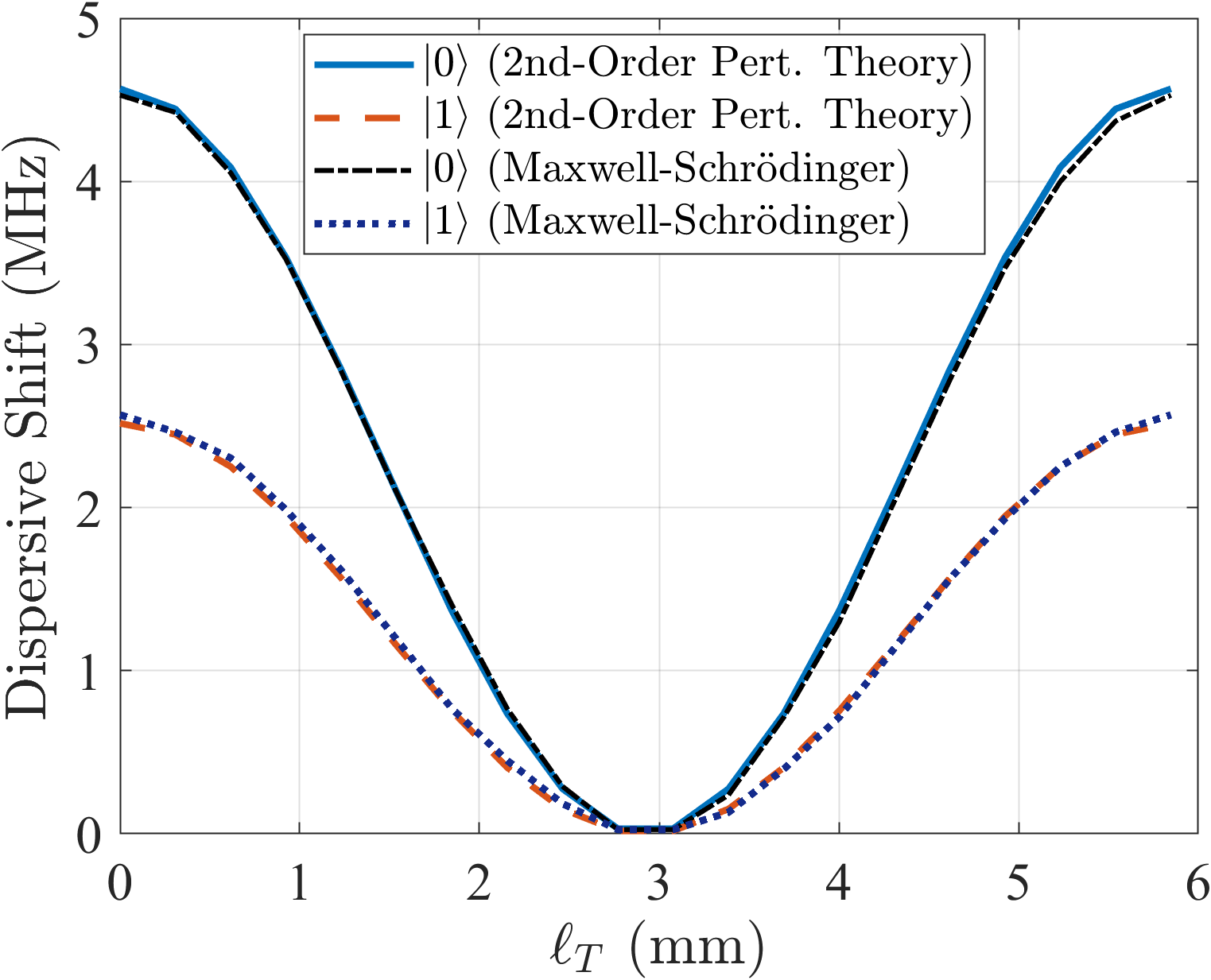}
	\caption{Comparison of analytical and numerical dispersive shifts of the transmission line resonator as a function of the transmon coupling location. Excellent agreement is achieved for every transmon location.}
	\label{fig:freq-sep}
\end{figure}

\subsection{Dispersive Regime Effects}
\label{subsec:qubit-measurement}
We now focus on modeling dispersive regime effects predicted from cavity quantum electrodynamics (QED) that are relevant to qubit state measurements to provide quantitative validation of the Maxwell-Schr\"{o}dinger method. The dispersive regime is achieved by significantly detuning the transmission line resonant frequency from the qubit transition frequencies \cite{krantz2019quantum}. These effects are typically derived using methods of cavity QED, which utilize a quantum treatment of the electromagnetic field \cite{koch2007charge,zhu2013circuit}. Here, we show that our semiclassical Maxwell-Schr\"{o}dinger method can also capture these effects. For all simulations, we use a reduced eigenstate discretization and consider the full Maxwell-Schr\"{o}dinger system with two-way coupling between the qubit and transmission line.

We begin by illustrating these effects for the transmon system shown in Fig. \ref{fig:transmon_readout_schematic}. For this setup, $E_J/E_C\! =\! 60$ and the first and second transition frequencies of the transmon are $4.60 \, \mathrm{GHz}$ and $4.35 \, \mathrm{GHz}$, respectively. Further, we set $N_\mathrm{eig} = 10$ to ensure accurate dynamics are computed. Accounting for the loading of the transmission line resonator, the first resonant frequency occurs at $\omega_r/(2\pi) = 5.971 \, \mathrm{GHz}$, which ensures the system is in the dispersive regime.

\begin{figure}[t!]
	\centering
	\includegraphics[width=0.95\linewidth]{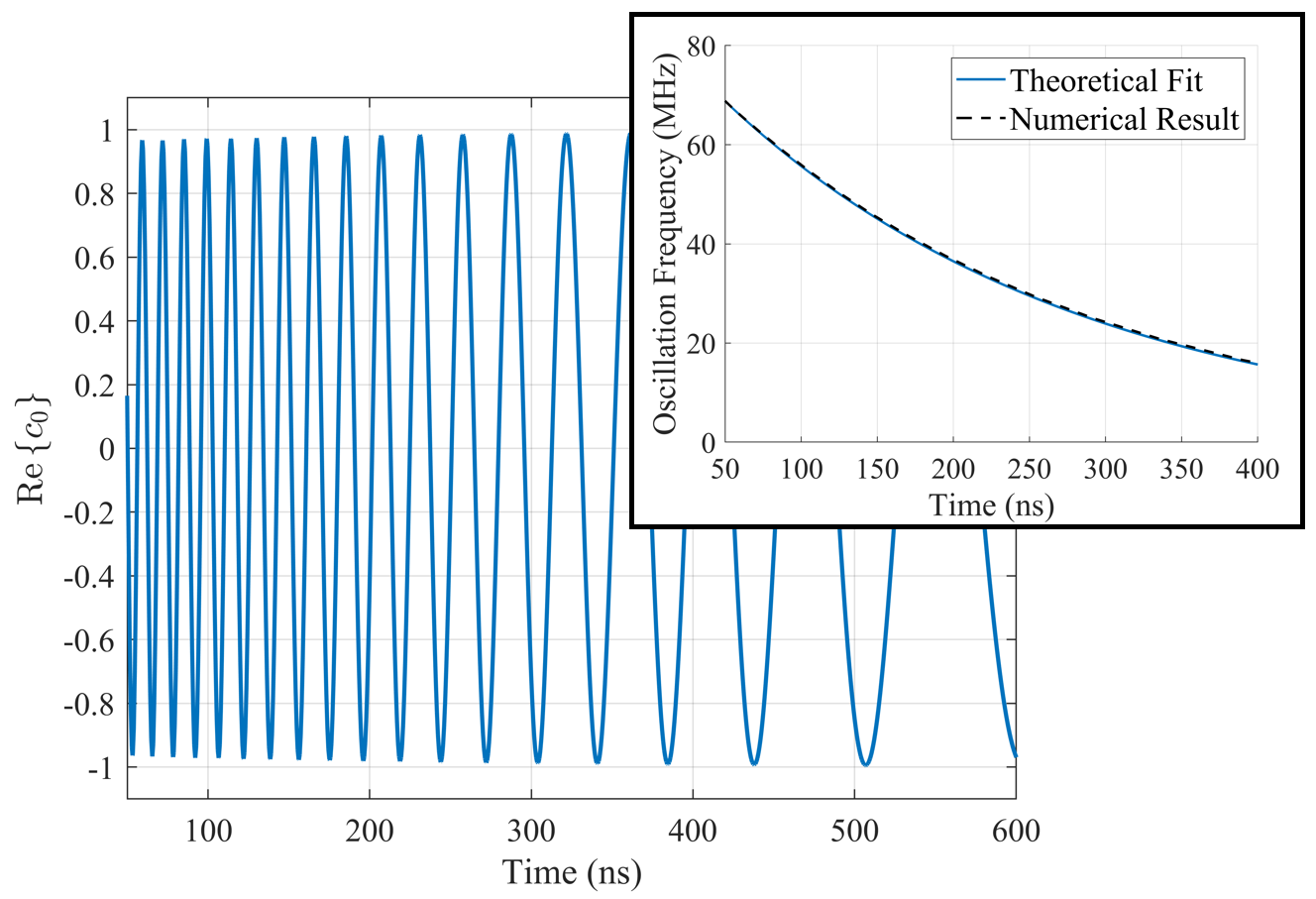}
	\caption{Oscillations in the excitation coefficient of $|0\rangle$ for $\ell_T = 1.847 \, \mathrm{mm}$. (Inset) Exponential fit using the theoretical resonator power decay rate yields excellent agreement with the oscillation frequencies observed in $|0\rangle$.}
	\label{fig:power-dependent-freq}
\end{figure}

A cavity QED analysis using standard 2nd-order perturbation theory of a transmon coupled to a transmission line resonator in the dispersive regime yields the following salient results \cite{zhu2013circuit}. Under a single resonator mode approximation, the coupling rate between the transmon and resonator is
\begin{align}
	g_{mm'} = \frac{2e\beta}{\hbar} \sqrt{\frac{\hbar \omega_r}{\ell_r C}} \cos\bigg( \frac{\pi}{\ell_r} \ell_T  \bigg) \langle m | \hat{n} | m' \rangle,
\end{align}
where $\ell_r$ and $\ell_T$ are defined in Fig. \ref{fig:transmon_readout_schematic}. Denoting the transition frequencies between state $|m\rangle$ and $|m'\rangle$ as $\omega_{mm'}$, the partial dispersive shifts are given as
\begin{align}
	\chi_{mm'} = \frac{|g_{mm'}|^2}{\omega_{mm'} - \omega_r}. 
\end{align}
The total dispersive shift of qubit level $m$ is given by
\begin{align}
	\chi_m = \sum_{m'}\big( \chi_{mm'} - \chi_{m'm} \big).
\end{align}
The frequency of the resonator exhibits a dispersive shift of $\chi_m$ if the measurement collapses the transmon state to $|m\rangle$. Finally, the dispersive regime also predicts that the oscillation frequency of a qubit state $|m\rangle$ will be modified by an amount proportional to $\chi_m$ times the number of photons in the resonator (which can be related to the power of the classical fields). 

\begin{figure*}[t!]
	\centering
	\includegraphics[width=0.9\linewidth]{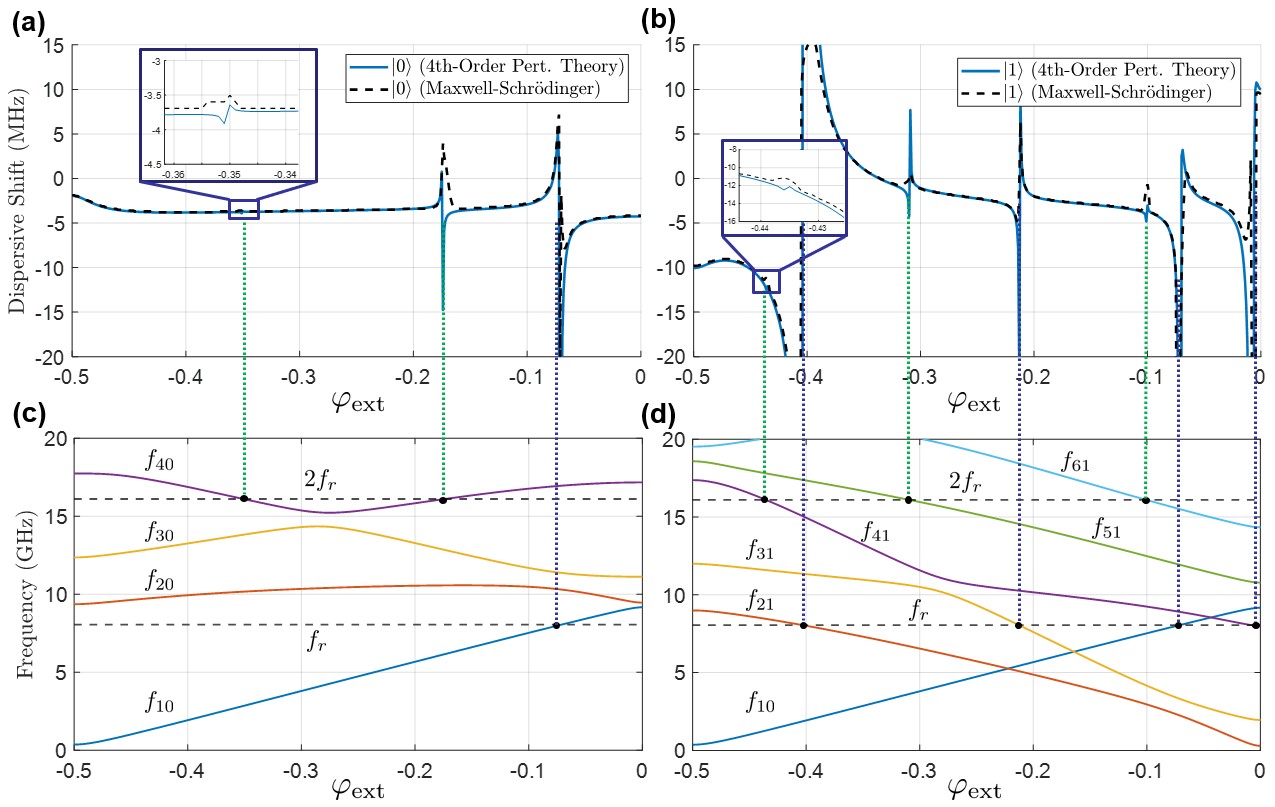}
	\caption{Dispersive shifts of the transmission line resonator computed when the qubit is initialized in the (a) ground or (b) first excited state as a function of an external magnetic flux characterized by $\varphi_\mathrm{ext}$. In (c) and (d), the relevant qubit transition frequencies are plotted, where $f_{ij}$ corresponds to a transition from state $j$ to state $i$. Also plotted are the transmission line resonance frequency $f_r$ and $2 f_r$. When the qubit transition frequencies intersect with the $f_r$ or $2 f_r$ lines a resonance in the dispersive shift is possible. Purple vertical dashed lines correspond to resonances that will always be present, while green vertical dashed lines correspond to higher-order nonlinear effects that can lead to vanishing resonances when the power in the resonator is low. As mentioned in the main text, the two methods do not need to quantitatively agree close to the resonances due to approximations in the theoretical model breaking down in this scenario.}
	\label{fig:dispersive-readout-results}
\end{figure*}

To numerically compute the dispersive shift, we run our Maxwell-Schr\"{o}dinger model when the transmon begins in the $|0\rangle$ or $|1\rangle$ state and compute the reflection coefficient. We extract the resonant peak locations and use these to determine $\chi_m$. We compare our numerical results to theoretical predictions of 2nd-order perturbation theory as a function of transmon coupling location in Fig. \ref{fig:freq-sep} and find excellent agreement. In computing the theoretical predictions, we also used a value of $N_\mathrm{eig}=10$ rather than the standard practice of only considering $N_\mathrm{eig} = 3$ for transmons. This changes the results by up to \textapprox4\%, but does noticeably improve the agreement between the Maxwell-Schr\"{o}dinger and perturbation theory models \cite{roth2023hybrid}.

We also compute the relative time-dependent oscillation frequency of the $|0\rangle$ state for $\ell_T = 1.847 \, \mathrm{mm}$, where the reference frequency is the expected free oscillation frequency in the absence of any microwave drive. The time-dependent oscillations are shown in Fig. \ref{fig:power-dependent-freq} with the extracted frequencies shown in the inset. We find that the oscillation frequency follows an exponential decay, as expected. We compute the Q-factor of the resonator using standard analytical approaches and find that the corresponding power decay rate exhibits an excellent fit to the decay rate of the oscillation frequency. 

Similar dispersive shifts also occur for fluxonium qubits, which we focus on demonstrating for the system shown in Fig. \ref{fig:fluxonium-device-schematic}. However, the strong nonlinearity of the fluxonium qubit greatly complicates the theoretical analysis, requiring the use of a 4th-order perturbation theory treatment to accurately describe dispersive regime effects \cite{zhu2013circuit}. Due to the complexity of this theoretical model, we do not review its details here for brevity. To compute the dispersive shifts with the Maxwell-Schr\"{o}dinger model, we follow a similar procedure as was described for the transmon system. We compare these results to the theoretical model of \cite{zhu2013circuit} as a function of the applied magnetic flux in Fig. \ref{fig:dispersive-readout-results}. We set $N_\mathrm{eig} = 10$ in both the Maxwell-Schr\"{o}dinger and theoretical models.

In each case, there are various resonances that occur in the dispersive shifts. From the theoretical model, it can be determined that these ``spikes'' should occur when a particular qubit transition frequency is resonant with the frequency of the transmission line resonator. In a standard 2nd-order perturbation theory treatment that is adequate for transmon qubits, these resonances only occur when a qubit transition frequency is exactly resonant with the transmission line resonator. However, the 4th-order perturbation theory treatment needed to describe fluxonium qubits shows that additional resonances can occur when a qubit transition frequency is resonant with twice the frequency of the transmission line resonator \cite{zhu2013circuit}. These higher-order effects are due to the stronger nonlinearity of the fluxonium qubit, and are also nonlinear in the sense that the amount of shift to the transmission line resonator frequency depends on the power in the resonator. For low resonator powers, some of these higher-order resonances can even vanish. For computing theoretical results, we used a single resonator power for all values of $\varphi_\mathrm{ext}$ for simplicity that was determined by finding the best fit with our Maxwell-Schr\"{o}dinger results. Finally, in comparing the behavior between the two models near the various resonances, it is important to note that they need not quantitatively agree because the approximations in the theoretical model can break down at these resonances. However, it is still important for all the relevant resonances to be observed between the two methods and at the correct locations. With this in mind, we see that the agreement between the theoretical model and our Maxwell-Schr\"{o}dinger method shown in Fig. \ref{fig:dispersive-readout-results} is excellent.

Our model can also compute dispersive shifts when the qubits are coupled to more complicated transmission line networks, which quickly becomes intractable using theoretical models or fully-quantum methods that require electromagnetic eigenmode decompositions. Further, our method can naturally be used to optimize the transient dynamics of the system to explore faster and higher fidelity qubit state measurement protocols, which is a significant need for emerging quantum computers to reach performance thresholds required to enable quantum error correction \cite{chen2021exponential,acharya2023suppressing}.

\section{Conclusion}
\label{sec:conclusion}
In this work, we presented the formulation of a self-consistent one-dimensional Maxwell-Schr\"{o}dinger method for analyzing the dynamics of a superconducting qubit capacitively coupled to a transmission line. We discussed two different discretization strategies that can be used to solve the Maxwell-Schr\"{o}dinger system of equations for general transmission line geometries. Numerical examples demonstrated the validity of the hybrid numerical method by comparing to established theoretical predictions. In the future, Maxwell-Schr\"{o}dinger methods can serve as a new tool for rapidly exploring broader design spaces to optimize control and measurement protocols for superconducting qubits.

Future work will consider extending this class of numerical method to include full-wave Maxwell solvers that will be important for characterizing emerging device architectures. Additionally, incorporating multi-qubit interactions is of significant interest to be able to model entangling gates and deleterious quantum crosstalk. Finally, developing open quantum system modeling methods in this Maxwell-Schr\"{o}dinger method will also be valuable to characterize the decoherence effects superconducting qubits are invariably subjected to in real-world devices.

\bibliographystyle{IEEEtran}
\bibliography{./paper_main_bib}

\end{document}